\documentclass[prl,aps,twocolumn]{revtex4}
\usepackage{rotate,bm,epsfig}
\newcommand{\be}{\begin{equation}}
\newcommand{\ee}{\end{equation}}
\newcommand{\ber}{\begin{eqnarray}}
\newcommand{\eer}{\end{eqnarray}}

\begin{document}
\title{Nanomagnetic toggle switching of vortex cores on the picosecond time scale}
\author{R. Hertel}
\affiliation{Institut f\"ur Festk\"orperforschung IFF-9 "Elektronische
   Eigenschaften", Forschungszentrum J\"ulich GmbH, D-52425 J\"ulich,
   Germany}
\author{M. F\"ahnle}
\affiliation{Max-Planck--Institut f\"ur Metallforschung,
  D-70569 Stuttgart, Germany}
\author{S. Gliga}
\affiliation{Institut f\"ur Festk\"orperforschung IFF-9 "Elektronische
   Eigenschaften", Forschungszentrum J\"ulich GmbH, D-52425 J\"ulich,
   Germany}
\author{C.M.~Schneider}
\affiliation{Institut f\"ur Festk\"orperforschung IFF-9 "Elektronische
   Eigenschaften", Forschungszentrum J\"ulich GmbH, D-52425 J\"ulich,
   Germany}
\date{\today}
\begin{abstract}
We present an ultrafast route for a controlled, toggle switching of
magnetic vortex cores with ultrashort unipolar magnetic field
pulses. The switching process is found to be largely insensitive to
extrinsic parameters, like sample size and shape, and it is faster
than any field-driven magnetization reversal process previously known
from micromagnetic theory. Micromagnetic simulations demonstrate that
the vortex core reversal is mediated by a rapid sequence of
vortex-antivortex pair-creation and annihilation sub-processes.
Specific combinations of field pulse strength and duration are
required to obtain a controlled vortex cores reversal. The
operational range of this reversal mechanism is summarized in a
switching diagram for a 200 nm Permalloy disk.  
\end{abstract}
\maketitle

Ferromagnetic materials in confined geometries typically form domain
structures that close the magnetic flux \cite{Hubert98}. In the center of such
flux-closure structures there is a region of only a few nanometers in
size known as a magnetic vortex, where the magnetization circulates
around a core \cite{Shinjo00,Wachowiak02,Miltat02b}. In the vortex
core, the magnetization points out  of the vortex plane, thereby
preventing a singularity of the exchange energy density. Owing to the
dramatic improvement in the spatial resolution of magnetic imaging
techniques, it has recently become possible to directly observe the
nanometric region of the vortex core \cite{Wachowiak02} and to study
vortex dynamics \cite{Choe04,Park03b,
Puzic05,Stoll_n06}.   

In analytical models for vortex dynamics \cite{Usov02}, 
the core is usually assumed to be a rigid magnetic structure that remains
unchanged when its position is shifted, {\em e.g.}, under the
influence of an external field. The high structural stability of
vortex cores results from the strong exchange interaction.
Nevertheless, recent studies have
shown that the internal magnetic structure of a vortex core can be
modified temporarily by magnetic fields applied in the film plane
\cite{Novosad05,Neudert05,Buchanan06}. Short of its 
annihilation with an antivortex \cite{Hertel06}, the most drastic
possible modification of a magnetic vortex structure is the reversal
of its perpendicular core, {\em i.e.}, the switching of the vortex
polarization.  

A field-driven switching of the vortex polarization has been shown to
be possible by applying a static magnetic field
\cite{Okuno02,Thiaville03} perpendicular to the vortex plane, thus
"crushing" the vortex core and 
re-establishing it in the opposite direction. This 
process requires field strengths of the order of 500 mT.
Such a large field value indicates that a high energy barrier
must be surmounted to switch a magnetic vortex core. This high barrier
ensures high thermal stability, which in combination with their
well-defined orientation and their extremely small size makes vortex
cores interesting candidates for binary data storage
\cite{Hollinger03}. The first demonstration that vortex cores   
can also be switched by low-amplitude in-plane magnetic fields
\cite{Zagorodny03} has been provided only very recently
\cite{Stoll_n06}.  
In the experiment of Ref.~\cite{Stoll_n06}, short oscillating magnetic
field pulses of low amplitude were used, tuned to the gyrotropic
resonance frequency of the system \cite{Choe04}. The gyrotropic
frequency depends on the particle size and 
shape and is typically in the order of a few 100\,MHz. 
A weak oscillating resonant magnetic field induces a rotation of 
the vortex on a stationary orbit. Exploiting the sense of rotation,
it was demonstrated that the vortex core reversal can be triggered
with weak sinusoidal field pulses of about 4\,ns duration and strength
of about 1.5 mT \cite{Stoll_n06}.      
\begin{figure}[ht]
\centerline{\epsfig{file=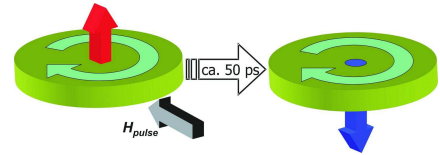,width=.65\linewidth}}
\caption{\label{scheme}Schematics of a field-pulse driven vortex core
switching.
The vortex core magnetization can be switched by just a short magnetic
field pulse applied in the film plane. This switching process requires
only 40-50 ps.} 
\end{figure}
 
\begin{figure*}[ht]
\centerline{\epsfig{file=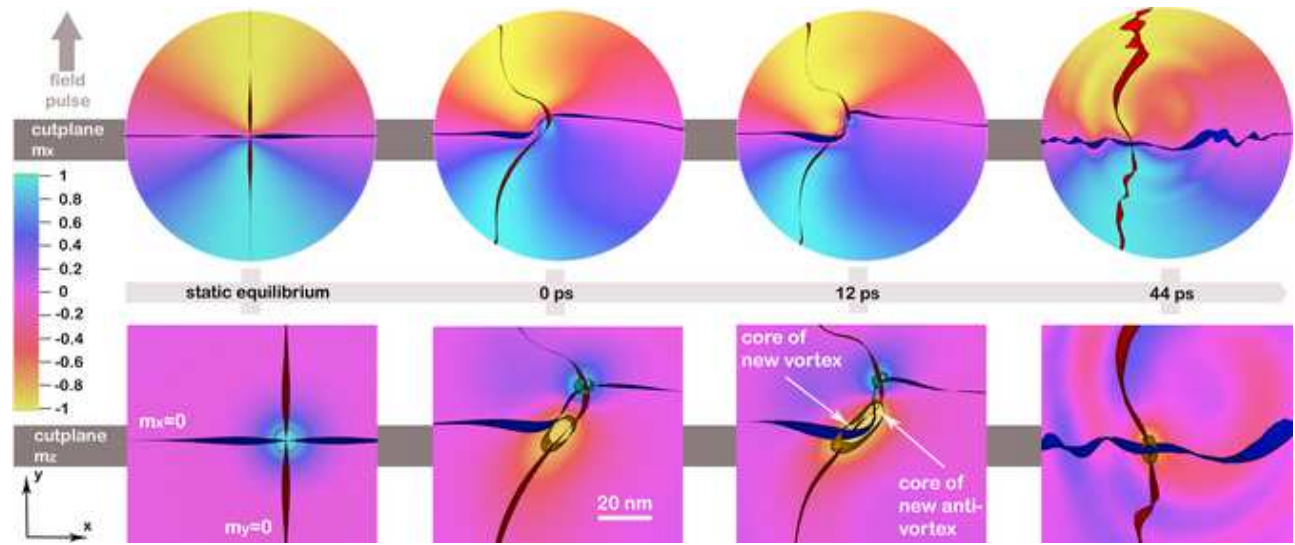,width=0.95\linewidth}}
\caption{\label{panel}Pair-creation mediated vortex core reversal in a
   Py disk of 100\,nm radius and 20\,nm thickness. A Gaussian
   field pulse is applied in the disk plane, parallel to the $y$ axis. 
   The top row shows the $x$ component of the magnetization, $m_x$, in
   the initial state, followed by the state of the sample at the pulse 
   maximum (``0 ps'') and then at two different times after the pulse maximum 
   has been reached. The blue and red ribbons represent the $m_x=0$ and $m_y=0$
   isosurfaces, respectively. The bottom row shows a magnification of the 
   region where these ribbons intersect, marking the cores of the
   original vortex, then at 12\,ps after the pulse maximum, of the
   newly created vortex anti-vortex pair and finally of the remaining
   vortex core. This core has opposite polarization with respect to
   the initial vortex core, indicated by the yellow color on the
   underlying cutplane representing the $z$-component of the
   magnetization, $m_z$. The green and orange cylindrical ribbons are
   the isosurfaces where $m_z=0.8$ and $m_z=-0.8$, respectively.} 
\end{figure*}

We have studied the dynamics of vortex core reversal with
 micromagnetic simulations using a fully three-dimensional
 finite-element algorithm based on the Landau-Lifshitz-Gilbert
 equation \cite{Hertel04}. Our simulations show that the
 time scale required for a vortex core reversal is not limited by
 the relatively slow gyrotropic resonance frequency: A vortex core
 reversal process can also be triggered by a non-resonant, unipolar,
 and very short field pulse (below 100 ps) of moderate strength
 ($\sim$80 mT) applied parallel to the film plane of, {\em e.g.}, a
 sub-micron sized Ni$_{81}$Fe$_{19}$ (Permalloy, Py) disk 
(Fig.~\ref{scheme}). In this study we also provide a detailed
description of the transformation undergone by the vortex during this
process: The core reversal occurs through a 
sequence consisting of a vortex-antivortex pair creation, followed by an
annihilation process, resulting in a final magnetic structure
of a single vortex with opposite polarization, in agreement
with the model suggested in Ref.~\cite{Stoll_n06}. We find that the
time required for the core reversal is of the order of 40\,ps
\footnote{The time required for pair creation depends on the pulse
  profile. However, the subsequent annihilation seems to
  occur within a constant time of about 20 ps.}. This is
a very high speed for a reversal process, which is, {\em e.g.}, about
five times faster than that involved in ultrafast precessional
switching mechanisms \cite{Gerrits02}.  

  A typical example of the vortex-antivortex pair creation  mediated
  core switch process is shown in Fig. 2. For the simulation of this
  Py disk (radius of 100\,nm and thickness of 20\,nm) we have used
  about 150.000 tetrahedral elements, corresponding to a cell size of
  3\,nm. The material parameters chosen for Py are: $A=13$\,pJ/m
  (exchange constant), $\mu_0M_{\rm s} =1.0$\,T ($M_{\rm s}$:
  saturation magnetization) and $K_{\rm u}$ = 0 
  ($K_{\rm u}$: magneto-crystalline anisotropy), corresponding to the
  usual room-temperature values \footnote{Thermal fluctuations have
  not been considered explicitely in this study.}. The Gilbert damping
  constant was set to $\alpha$=0.01. An 80\,mT Gaussian shaped field
  pulse of a duration of 60\,ps is applied in the   
  plane of the Py disk, which is initially in a  
  symmetric vortex state. The  microscopic processes  
  leading to the core reversal can be clearly identified by
  highlighting  the $m_x=0$ and $m_y=0$ isosurfaces \cite{Hertel06},
  the intersection of which determines the exact position of the
  vortex core.  Before an external field is applied, these
  $m_x=M_x/M_{\rm s}=0$ and $m_y=M_y/M_{\rm s}=0$ isosurfaces
  appear as straight ribbons (oriented parallel to the $x$ and $y$ 
  axis, respectively) crossing each other perpendicularly at the center
  of the vortex core. As the field pulse perturbs the system, the vortex
  shifts away from its original position and the formerly circular
  arrangement of the magnetization around the core is stretched in one
  direction, resulting in bent isosurfaces. A few picoseconds after
  the peak value of the pulse is reached, the isosurfaces are bent
  strongly enough to form two additional intersections. These 
  intersections mark the creation of a vortex-antivortex pair
  \cite{Hertel06}. 
  The magnetization direction in the core of the new  
  vortex is opposite to the polarization of the original core.  
\begin{figure}[ht]
\centerline{\epsfig{file=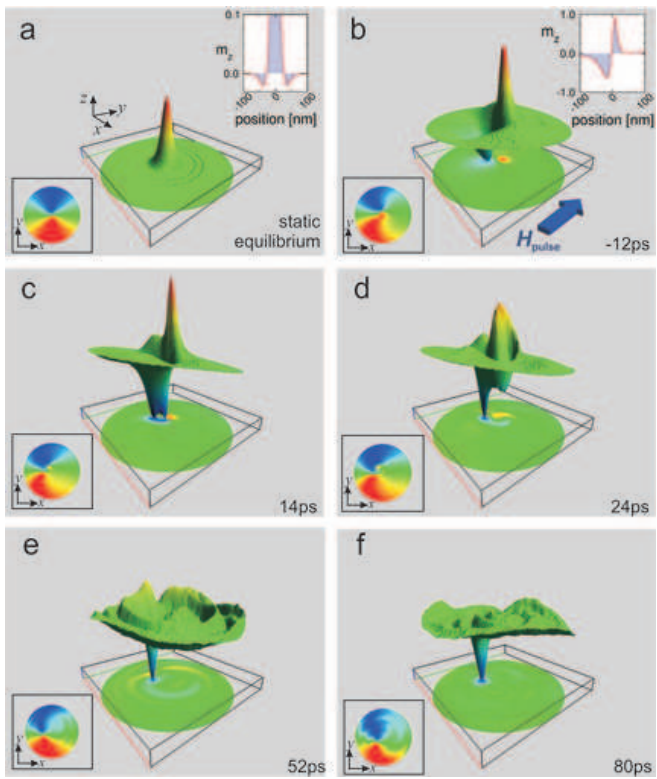, width=\linewidth}}
\caption{\label{topo}Topography of $m_z$ during the core switching
   process. (a): Static equilibrium. The cutline on the upper right
   shows the small negative dip in $m_z$ near the core. (b): Formation
   of a large negative dip of $m_z$.  The cutline 
   goes along the diameter through the vortex core and the adjacent
   dip. (c): Pair creation leads to two separate points with
   $m_z$=$-1$. (d): The original vortex core annihilates with the new
   antivortex. (e): Spin waves are emitted after the 
   annihilation. Finally, a vortex core with opposite
   polarization remains (f). The insets on the lower left show 
   the $x$-component of the magnetization (red: $m_x=1$, blue:
   $m_x=-1$). 
The times are relative to the pulse maximum.}  
\end{figure}

Once a pair is created, the antivortex quickly moves towards the
original vortex and, through a rapid process, they annihilate each
other. The details of the magnetization dynamics of such annihilation
processes, which include the formation and the propagation of a
micromagnetic singularity (Bloch point), have been reported in
\cite{Hertel06}. It has been shown that this vortex-antivortex
annihilation process is connected with a sudden generation of spin
waves \cite{Lee06}. After the sequence of pair creation and
annihilation  processes, the magnetic structure is again in a vortex
state, but with opposite polarization with respect to the original
vortex. The process does not affect the in-plane sense of rotation of
the magnetization, hence resulting in a change of the vortex
handedness \cite{Choe04}.   

The change of the perpendicular core magnetization, while not
explicitly shown in Fig. 2, can be seen in Fig. 3. The reversed core
magnetization is clearly obtained by the creation of a new oppositely
magnetized area, which evolves into a vortex-antivortex
structure. Subsequently, the original core annihilates with the new
antivortex. The starting point for the formation of the out-of-plane
component of the new vortex is a "dip" in the perpendicular component
of the magnetization, $m_z$, that forms close to the vortex core
\cite{Novosad05,Kasai06} as the vortex is distorted, cf.~Fig.3b. 
The formation of this out-of-plane component is due to the tendency
of the system to reduce the exchange energy connected with the strong
inhomogeneity of the in-plane 
magnetic structure: as the $m_x$ = 0 and the $m_y$=0 isosurfaces
approach each other, the distance over which the magnetization changes
its in-plane direction by 90$^\circ$ is reduced to only a few
nanometers. The system circumvents the formation of such a strongly
inhomogeneous structure by rotating the magnetization out of the
plane.  While the exchange field  
is responsible for the {\em formation} of a pronounced out-of-plane
component in a distorted vortex structure, 
it is the magnetostatic field of the vortex core that is decisive for
the {\em direction} into which this out-of-plane dip develops. In the
close vicinity of the vortex core, there is a strong dipolar field
originating from the core. This field provides a bias into the
opposite direction of the vortex core magnetization. As shown on the
cutline in Fig. 3a, the dipolar field of the core affects the magnetic
structure on a ring around the vortex core, leading to a negative
"halo" of the $m_z$ component around the core
\cite{Ha03}.
 The simulations yield that the vortex core gives rise to a negative 
perpendicular field component of about 200 mT in the region where the 
dip develops \footnote{The importance of the dipolar field in
  selecting the polarization of the vortex-antivortex pair has
  been shown by applying various out-of-plane field pulses
  parallel to the original vortex core polarization in addition to
  an in-plane pulse. The pair-creation process that is usually induced
  by the in-plane pulse is suppressed if the out-of plane pulse
  exceeds a magnitude of about 200 mT, corresponding to 
  the value of the dipolar field strength.}

The simulations show that the pair-creation mediated
vortex core reversal mechanism is insensitive to variations in
particle shape or size. The same
reversal process occurs as well in elliptical and square sub-micron
sized magnetic thin-film elements. The process also does not seem to
depend on the value of the damping coefficient, $\alpha$, used in the
Landau-Lifshitz-Gilbert equation  
\footnote{See EPAPS Document No. XXX for three movies. movie1.avi details the 
isosurface representation. movie2.avi shows the vortex core reversal
in the disk sample presented in Fig. 2 and movie3.avi the same process
observed in a $300$x$125$x$20$ nm Py elliptical element where
$\alpha=0.5$. The visualization was done using GMV:
http://www-xdiv.lanl.gov/XCM/GMVHome.html. For more information on
EPAPS, see http://www.aip.org/pubservs/epaps.html. }. 
Only the characteristics of the
required field pulse depend on the individual sample
properties. These results suggest that the field-pulse induced 
generation and the subsequent annihilation of magnetic
vortex-antivortex pairs is a general property of magnetic vortices.
In view of this general nature of the process, we assume
 that the pair-creation mediated core reversal could also be initiated
 by electric current pulses \cite{Kasai06} or by laser pulses
 \cite{Kimel05}.

\begin{figure}
\centerline{\epsfig{file=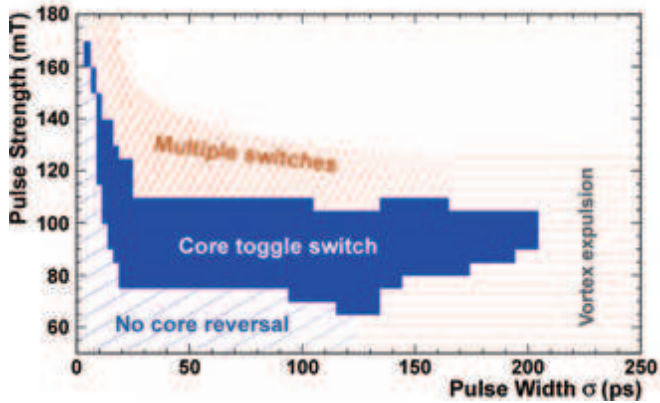,width=\linewidth}}
\caption{\label{phasediag}Switching diagram for the pulse parameters
   leading to a toggle switch of the vortex core in a 100 nm radius
   and 20 nm thick  Py disk. The pulse duration is quantified  
   by its width, $\sigma$, and the strength by its maximum value. The
   cell size was 3\,nm, varied by $\pm$0.7\,nm at 
   80\,mT / 20\,ps (diagram edge), giving a variation in the field
   required to switch the core of +5.0\,mT for 2.3\,nm cells and
   -5.6\, mT for 3.7\,nm cells.}
\end{figure}

We found that the field-pulse induced core switch in our disk-shaped
sample occurs for well-defined combinations of the applied 
pulse's duration and strength. The influence of these two parameters
is shown in Fig.~\ref{phasediag} on the example of a 20 nm thick, 200
nm large Py nanodisk. This diagram was obtained by varying the pulse
width $\sigma$ in steps of 2.5\,ps for $t = 0$\,ps to $t = 20$\,ps and
in steps of 10\,ps thereafter. The pulse maximum was varied in
increments of 5\,mT. The precise position of the boundaries in the
diagram is difficult to determine because they (weakly) depend on the
size of the  discretization cells. These variations in the critical
field result from the mesh-dependence of the Bloch point nucleation
\cite{Thiaville03}. The operating field range is relatively narrow: 
While too low fields do not cause the core to switch, too high
fields give rise to sequences of multiple pair creations and
annihilations.
It is observed that by extending the pulse duration, switching can be 
accomplished with pulses of 65\,mT, while by increasing the applied
field strength, it is possible to obtain switching with pulses only
5\,ps long. Ultimately, too strong and too long pulses lead to a 
temporary expulsion of the vortex from the sample.

In conclusion we presented a detailed description of an ultrafast
process in nanomagnetism for switching the core of magnetic
vortices. Although this reversal mode involves a series of complex
processes on the nanometer scale - the creation of a vortex-antivortex
pair, a subsequent annihilation process - 
the chain of events only requires a short  
field pulse of suitable shape to be initiated. The ultrafast speed of
this process is perhaps equally astounding as the finding that such a
long sequence of subprocesses develops almost automatically.  

The pair-creation mediated vortex core reversal is comparable in scope
with the precessional switching mechanism \cite{Gerrits02}. However,
the pair-creation and annihilation processes are
driven by the exchange field, while the precessional switching
exploits the demagnetizing field. The magnitude of the exchange field
is in the order of 100\,T, which is about 100 times larger than the
demagnetizing field. This explains the considerably higher speed of
the core reversal, which is almost an order of magnitude faster. A
further advantage of the vortex core reversal mechanism lies in the
simplicity of the required sample and magnetic structure: all that is
needed is a magnetic vortex, which is a structure that forms naturally
in sub-micron sized magnetic disks \cite{Cowburn99}. 
The core reversal process described here therefore represents a
significant leap towards smaller length scales and shorter time
scales. 

We thank H.~Stoll (MPI Stuttgart) and coworkers for bringing
the vortex switching problem to our attention.  

\end{document}